\documentstyle[fullpage,12pt, doublespace, rangecite]{article}

\begin{document}

\title{Observations of Quantum Dynamics
by Solution-State NMR Spectroscopy\\}


\author{Marco Pravia$^{\dagger}$,  Evan Fortunato$^{\dagger}$,  Yaakov Weinstein$^{\ddagger}$, Mark D. Price$^{\P}$,\\ 
Grum Teklemariam$^{\S}$, Richard J. Nelson$^{\ddagger}$, Yehuda Sharf$^{\dagger}$, \\
Shyamal Somaroo$^{\S\S}$, C.H. Tseng$^{\S\S\#}$, Timothy F. Havel$^{\S\S}$, David G. Cory$^{\dagger*}$\\
$^{\dagger}${\normalsize Department of Nuclear Engineering, Massachusetts Institute of Technology, Cambridge, MA 02139}\\
$^{\ddagger}${\normalsize Department of Mechanical Engineering, Massachusetts Institute of Technology, Cambridge, MA 02139}\\
$^{\P}${\normalsize Health Sciences and Technology, Massachusetts Institute of Technology, Cambridge, MA 02139}\\
$^{\S}${\normalsize Department of Physics, Massachusetts Institute of Technology, Cambridge, MA 02139}\\
$^{\S\S}${\normalsize BCMP Harvard Medical School , 240 Longwood Avenue, Boston MA 02115}\\
$^{\#}${\normalsize Center for Astrophysics, Harvard Smithsonian, Cambridge MA 02138}\\
\\
$(^{*})$ {\normalsize Author to whom correspondence should be sent}}

\date{\today}

\headheight 0.3in
\headsep 0.2in
\oddsidemargin -0.2in
\evensidemargin -0.2in
\textheight 9.2in
\textwidth 6.55in
\topmargin -.5in

\newcommand{\ket}[1]{$\vert${#1}$\rangle$}
\newcommand{\mket}[1]{\vert{#1}\rangle}
\newcommand{\mbra}[1]{\langle{#1}\vert}
\newcommand{\tfrac}[2]{{\textstyle\frac{#1}{#2}}}

\def\up{|\uparrow\,\rangle}
\def\dn{|\downarrow\,\rangle}
\def\upd{\langle\, \uparrow |}
\def\dnd{\langle\, \downarrow |}
\def\beqn{\begin{equation}}
\def\eeqn{\end{equation}}
\def\beqnar{\begin{eqnarray}}
\def\eeqnar{\end{eqnarray}}
\oddsidemargin -0.2in
\evensidemargin -0.2in
\newcommand{\mb}[1]{\mbox{\boldmath{$#1$}}}
\def\ba{\begin{array}}
\def\ea{\end{array}}
\newcommand{\wdg}{\! \wedge \!}
\newcommand{\crs}{\! \times \!}
\newcommand{\scp}{\! \ast \!}
\newcommand{\dt}{\! \cdot \!}
\newcommand{\etal}{{\em et al. }}
\newcommand{\eqn}[1]{(\ref{#1})}

\maketitle

\begin{abstract}

NMR is emerging as a valuable testbed for the investigation of foundational 
questions in quantum mechanics.  The present paper outlines the
preparation of a class of mixed states, called 
pseudo-pure states, that emulate pure quantum states in the 
highly mixed environment typically used to describe solution-state 
NMR samples.  It also describes the NMR observation of spinor 
behavior in spin 1/2 nuclei, the simulation of wave function 
collapse using a magnetic field gradient, the creation of entangled 
(or Bell) pseudo-pure states, and a brief discussion of quantum 
computing logic gates, including the Quantum Fourier Transform.
These experiments show that liquid-state NMR can be used to demonstrate
quantum dynamics at a level suitable for laboratory exercises.
  
\end{abstract}

\section{Introduction}

The fundamental physics of NMR is again, 50 years after its discovery, the 
subject of much discussion.  The impetus behind this recent
interest is the dramatic potential of quantum information processing
(QIP) \cite{Steane}, particularly quantum computing, along with the 
realization that liquid-state NMR provides an experimentally accessible 
testbed for developing and demonstrating these new ideas \cite{coryorig,cory1,coryphysica,gershen,knill97,jonesjchem,chuangseth,chuangkubi,jonesmosca,jonesscience,cory2}.

Most descriptions of quantum information processors have focused on the 
preparation, manipulation, and measurement of a single quantum system in a 
pure state.  The applicability of NMR to QIP is somewhat surprising 
because, at finite temperatures, the spins constitute a highly mixed 
state, as opposed to the preferred pure state.  However, NMR technology
applied to the mixed state ensemble of spins (the liquid sample) does offer 
several advantages.  Decoherence, which plays a detrimental role in the storage of 
quantum information, is conveniently long (on the order of seconds) in a typical 
solution sample, and it acts on the system by attenuating the elements of the density matrix and rarely mixes them.  NMR spectrometers allow for precise control of the spin system via the application of arbitrary sequences of RF excitations, permitting the
implementation of unitary transformations on the spins.  Effective non-unitary 
transformations are also possible using magnetic field gradients.  The
gradient produces a spatially varying phase throughout the sample, and since 
the detection over the sample is essentially a sum over all the spins, 
phase cancellations from spins in distinct positions occur.  
These characteristics of NMR enable the creation of a class of mixed states, called pseudo-pure states, which transform 
identically to a quantum system in a pure state\cite{cory1}. 

NMR does have several noteworthy disadvantages.  A single density matrix cannot be associated
with a unique microscopic picture of the sample, and the close proximity of
the spins prevents the study of non-local effects.  Additionally, the preparation of pseudo-pure
states from the high temperature equilibrium state in solution NMR entails an exponential
loss in polarization.  \cite{warren}

In this paper, we review the results of a number of simple NMR experiments 
demonstrating interesting quantum dynamics.  The experiments illustrate spinor 
behavior under rotations, the creation and validation of pseudo-pure states,
their transformation into ``entangled'' states, and the simulation of wave 
function collapse via gradients.  Additionally, the implementations of basic quantum logic gates are described, along with the Quantum Fourier Transform.

\section{The Spin System}

The experiments were performed on the two-spin heteronuclear spin system,
$^{13}$C-labeled chloroform ($^{13}$CHCl$_3$), thereby eliminating the 
use of shaped RF pulses.  The $^{13}$C (I) and the $^1$H (S) nuclei interact
via weak scalar coupling, and the Hamiltonian for this system is written as 
  \begin{equation}
    {\cal{H}} =\omega_{I}I_{z}+\omega_{S}S_{z} 
    +2\pi J I_{z}S_{z}, 
  \end{equation} 
where $\omega_{I}$ and $\omega_{S}$ are the Larmor frequencies of the $^{13}$C 
and $^1$H spins respectively and $J<<\vert\omega_{I}-\omega_{S} \vert$ is 
the scalar coupling constant.  

In the standard model of quantum computation, the quantum system is
described by a pure state.  However, liquid-state 
NMR samples at room temperature are in highly mixed states, requiring 
the state of the system to be described by the density operator.  
In a liquid sample, the inter-molecular interactions are, for most practical
purposes, averaged to zero so that only interactions within a molecule are 
observable; in other words, the sample can be thought of as an ensemble of
quantum processors, each permitting quantum coherence within but not between
molecules.  For the purposes of this paper, the large density matrix of size
$2^N \times 2^N$, where N is the number of spins in the 
sample, may be replaced by a much smaller density matrix of 
size $2^n \times 2^n$, where $n$ is the number of distinguishable 
spin-$\tfrac{1}{2}$ nuclei in the molecule.  In the high temperature regime 
($\epsilon=\frac{\hbar\gamma_IB_o}{2kT} \sim {\cal{O}}(10^{-6})$) the 
equilibrium density operator for the ensemble is
\beqn
 \rho=\tfrac{e^{-{\cal H}/kT}}{Z}\approx \tfrac{1}{4}{\bf 1}+\tfrac{1}{4}\epsilon\rho_{dev}=
  \tfrac{1}{4}{\bf 1}+\tfrac{1}{4}\epsilon \left(I_{z} + \frac{\gamma_S}{\gamma_I}S_{z} \right),
\eeqn
where the relative value of the gyromagnetic ratios is $\gamma_S/\gamma_I \sim 4$.  

From the above, it is clear that at room temperature a spin system cannot be prepared in a pure state.  However, it is possible to prepare a pseudo-pure state that transforms like a pure state.  Also, notice that since the identity part of the density operator is invariant under unitary transformations, it is the deviation part of the density operator, that holds the information on the spin dynamics.  Henceforth in this paper, the deviation density matrix will be simply referred to as the density matrix.  The density operator is often written in the product operator basis formed by the direct product of individual spin operators\cite{productop,somaroo}.  The product operator technique is used throughout this paper to express the dynamics of the spin system.  Furthermore, if  $n$ spins are coupled to one another, any arbitrary unitary operation can be composed from a series of RF pulses, chemical shift evolution and scalar coupling evolutions.   \cite{coryphysica,BBCDMSSSW:95}



\section{Preparation of Pseudo-Pure States}
Before describing the creation of the pseudo-pure state, it is
convenient to begin with a system of equal spin populations.  This is 
achieved by applying the pulse sequence
\begin{equation}
  \left[ \frac{\pi}{2} \right]^{I,S} _{x} \rightarrow
  \left( \frac{1}{4J} \right) \rightarrow
  \left[ \frac{\pi}{2} \right]^{I,S} _{y}\rightarrow
  \left( \frac{1}{4J} \right) \rightarrow
  \left[ \frac{\pi}{2} \right]^{I,S} _{-x} \rightarrow
  \left[ grad(z) \right],
\end{equation}
to the equilibrium density matrix, resulting in
\beqn 
\frac{1}{4}{\bf 1} + \frac{\epsilon}{4}\left(1+\tfrac{\gamma_S}{\gamma_I}\right)(I_z + S_z),
\label{eq:bal}
\eeqn
which has a balanced spin population. Because the eigenvalue structure of this density matrix is different from that of thermal equilibrium, there is no unitary transformation which could transform one to the other.  The non-unitary gradient (where the non-unitarity refers to the spatial average over the phases created by the gradient) at the end of the above pulse sequence makes this transformation possible.  Figure \ref{ppprepfig} shows a spectrum obtained after applying this sequence.

Since the identity part of the equalized density matrix is unaffected by unitary transformations and undetectable in NMR, only the deviation density matrix,
\begin{equation}
I_{z}+S_{z}\quad=\quad
\ba{rl}
 & \ba{cccc}| 0^{\tiny I} 0^{\tiny S}\rangle &  |0^{\tiny I}1^{\tiny S}\rangle & |1^{\tiny I}0^{\tiny S}\rangle & | 1^{\tiny I}1^{\tiny S}\rangle \ea \\

\ba{c}
\langle 0^{\tiny I}0^{\tiny S}| \\
\langle 0^{\tiny I}1^{\tiny S}| \\
\langle 1^{\tiny I}0^{\tiny S}| \\
\langle 1^{\tiny I}1^{\tiny S}| \\
\ea
 &
{
\left(
    \begin{array}{p{11.5mm}p{11.5mm}p{11.5mm}p{11.5mm}}
      1&0&0& 0\\
      0&0&0& 0\\
      0&0&0& 0\\
      0&0&0&-1
    \end{array}
\right)},
\ea
  \end{equation}
which represents the excess magnetization aligned with the external magnetic field, is of interest.  The above matrix representation has been made in the eigenbasis of the unperturbed Hamiltonian, and here the rows and columns have been labeled explicitly to avoid ambiguity.  In the subsequent matrix expressions, the labels will be dropped.

QIP requires the ability to create and manipulate pure states.  NMR systems, however, are in a highly mixed state at thermal equilibrium.  While single spin manipulation is not feasible in NMR, Cory et. al. \cite{coryorig,cory1,gershen} have developed a technique by which the equilibrium state is turned into a pseudo-pure state.  Such a state can be shown to transform identically to a true pure state as follows:  according to the rules of quantum mechanics, a unitary transformation ${\cal U}$ maps the density matrix $\rho$ to $\rho'={\cal U} \rho {\cal U}^{\dag}$.  Thus an $N$-spin density matrix of the form $\rho=({\bf{1}}+\mket{\psi} \mbra{\psi})/2^N$ is mapped to 
\beqn
\frac{\bf{1}+({\cal U} \mket{\psi})({\cal U}\mket{\psi})^{\dag} }{2^N}. 
\eeqn
This shows that the underlying spinor $\mket{\psi}$ is transformed one-sidedly by ${\cal U}$ just as a spinor which describes a pure state would be. 

After equalizing the spin population from the thermal equilibrium state (eq. (5)), the application of
\begin{equation}
  \left[ \frac{\pi}{4} \right]^{I,S} _{x} \rightarrow
  \left( \frac{1}{2J} \right) \rightarrow
  \left[ \frac{\pi}{6} \right]^{I,S} _{y}\rightarrow
  \left[ grad(z) \right]
\end{equation}
results in the pseudo-pure state (neglecting the initial identity component)
\beqn 
  \sqrt{\frac{3}{32}}{\bf 1}+\sqrt{\frac{3}{8}}\left(I_{z}+S_{z}+2I_{z}S_{z}\right)=
    \sqrt{\frac{3}{2}}
    \left(
    \begin{array}{rrrr}
      1&0&0& 0\\
      0&0&0& 0\\
      0&0&0& 0\\
      0&0&0& 0
    \end{array}\right).
\eeqn

Figure \ref{pptomofig} shows a series of spectra confirming the preparation of a pseudo-pure state.

\section{Spinor Behavior}

Particles of half-integral spin have the curious property that when rotated
by $2\pi$, their wave functions change sign while a $4\pi$ rotation returns 
their phase factors to their original value.  The change in the sign of the
wavefunction is not observable for a single particle, but it can be seen
through an interference effect with a second ``reference spin.''  Spinor 
behavior, as this effect is called, was first experimentally measured
using neutron interferometry \cite{neutron1,neutron2} and later using 
NMR interferometry \cite{vaughn}. 

The following simple experiment describes how the spinor behavior can be
seen in chloroform, where the spinor behavior of $^{13}$C is correlated with the $^1$H nuclei as a multiplicative phase factor.  Consider the unitary transformation
\beqn
{\cal U}=
\left(
    \begin{array}{cccc}
      1&0&0&0\\
      0&\cos\left(\tfrac{\phi}{2}\right)&0&-\sin\left(\tfrac{\phi}{2}\right)\\
      0&0&1&0\\
      0&\sin\left(\tfrac{\phi}{2}\right)&0&\cos\left(\tfrac{\phi}{2}\right)
    \end{array}\right)=
e^{-i\phi I_y (\tfrac{1}{2} - S_z)}.
\eeqn
As explained in section 6, this can be viewed as a rotation by $\phi$ of the $^{13}$C conditional on the $^1$H being in the down state.  This can be implemented via the pulse sequence
\beqn
  \left[ \frac{\phi}{2} \right] _{y}^I  \rightarrow
  \left[ \frac{\pi}{2} \right] _{x}^I \rightarrow
  \left[ \frac{\phi}{2\pi J} \right] \rightarrow
  \left[ \frac{\pi}{2} \right] _{-x}^I.
\eeqn
Application of this pulse sequence to the state $2I_zS_x$, where the spinor behavior of the I-spin is revealed by its correlation to the S-spin, results in
\beqn
2\cos(\phi/2)I_zS_x + 2\sin(\phi/2)I_xS_x.
\eeqn
It can be clearly seen that when $\phi=2\pi$ the initial state gains
a minus sign, but when $\phi=4\pi$ the state returns to its initial value. 
The state $2I_zS_x$ is made observable under the evolution of the internal hamiltonian previously defined and can be created from the equalized equilibrium state (eq. \ref{eq:bal}) using the sequence
\beqn
  \left[ \frac{\pi}{2} \right] _{x}^I \rightarrow
  \left[ grad(z) \right] \rightarrow
  \left[ \frac{\pi}{2} \right] _{x}^S \rightarrow
  \left( \frac{1}{2J} \right).
\eeqn
Figure \ref{spinorfig} shows the spectra for several values of $\phi=0,2\pi$, and $4\pi$.

\section{Entangled States}

The Einstein-Podolski-Rosen (EPR) \cite{epr:orig,b:qt} paradox, concerning the 
spatial correlations of two entangled quantum systems, is perhaps the most 
famous example of quantum dynamics that is incompatible with a classical view.  An entangled state is one that cannot be factored into the product of the individual particle wavefunctions.  As a result, the state of one particle is necessarily correlated with the state of the other, and these correlations differ from those allowed by classical mechanics.  Entanglement in quantum mechanics is normally raised to explore aspects of non-local effects and hidden variable theories.  Due to the close proximity of nuclear spins and the fact that the ensemble is in a highly mixed state, the NMR measurements discussed below do not address these issues.  Nevertheless, we can use the ability of liquid state NMR to simulate strong measurement to show that the behavior of an entangled state is inconsistent with a simple classical picture.

The entangled state $\mket{\psi}=\frac{1}{\sqrt{2}}(\mket{00}+\mket{11})$, otherwise known as a Bell state, is given by the density matrix
\beqn
\rho_{\rm Bell} =\tfrac{1}{2}\left(\tfrac{1}{2}{\bf 1} + 2I_zS_z + 2I_xS_x -2I_yS_y \right). 
\eeqn 
The above state can be prepared directly from the pseudo-pure ground state $|00\rangle$ by the transformation
\beqn
{\cal U} \equiv e^{-i I_x S_y \pi}
\eeqn
which is implemented by the pulse sequence
\beqn  
  \left[ \frac{\pi}{2} \right] ^S _{-x}  \rightarrow
  \left[ \frac{\pi}{2} \right] ^I _{y}   \rightarrow
  \left( \frac{1}{2J} \right)   \rightarrow
  \left[ \frac{\pi}{2} \right] ^I _{-y}  \rightarrow
  \left[ \frac{\pi}{2} \right] ^S _{x}.
\eeqn
Readout pulses can then be used to verify the creation of this Bell state, as shown in Fig \ref{belltomofig}. 

One of the advantages of working with an ensemble is that we can introduce a pseudo-random phase variation accross the sample to simulate the decoherence that accompanies strong measurement.  A pseudo-random phase variation in a given basis can be achieved by rotating the preferred axis to the z-axis and then applying a magnetic field gradient followed by the inverse rotation.  This leads to the pulse sequence
\beqn 
  \left[ \frac{\pi}{2} \right] ^I _{y} \rightarrow
  \left[ grad(z) \right] \rightarrow
  \left[ \pi \right] ^S _{y}  \rightarrow
  \left[ grad(z) \right] \rightarrow
  \left[ \frac{\pi}{2} \right] ^I _{-y}.
\label{xmeasure}
\eeqn
It can be shown that such a measurement also ``collapses'' the $S$ spin along this direction. Thus, half the magnetization is along the +x-axis and the other half is along the -x-axis leaving zero magnetization in the $y$--$z$ plane. This is verified in our experiment by applying a series of readout pulses to confirm the creation of the $2I_xS_x$ state which corresponds to ``collapsing'' the pseudo-pure Bell state along the x-axis.  The experimental results are shown in Fig \ref{strongmeasure}.

An incoherent mixture of entangled states is easily generated by the pulse sequence
  \begin{equation}
    \left[ \frac{\pi}{2} \right] _{90^{\circ}}^{S}\rightarrow
    \left( \frac{1}{2J} \right) \rightarrow
    \left[ \frac{\pi}{2} \right] _{135^{\circ}}^{I}\rightarrow
    \left( \frac{1}{2J} \right) \rightarrow
    \left[ \frac{\pi}{2} \right] _{90^{\circ}}^{S}
  \end{equation}
applied to $\rho_{eq}$ (Eq. ~\ref{eq:bal}), yielding the reduced density matrix
  \begin{equation}
    \rho_f = \left(\begin{array}{cccc}
                0&          0&0&\frac{-1-i}{\sqrt{2}}\\
                0&          0&0&           0         \\
                0&          0&0&           0         \\
      \frac{-1+i}{\sqrt{2}}&0&0&           0
    \end{array}\right).
  \end{equation}

Suppose one wishes to measure the polarization of spin $I$ along the $x$--axis
 and spin $S$ along the $z$--axis.
One possibility is to use selective RF pulses to rotate the desired axis ($x$ in this case) to the
  $z$--axis, apply a $z$-gradient, and then rotate back to the $x$--$y$ plane to observe the induction signal as in Eq.~\ref{xmeasure}.
Alternatively, one could rotate the desired measurement axis of one of the
  spins to the $z$--axis, rotate the other spin to the $x$--$y$ plane
  and then spin-lock the sample on resonance.
In this latter case the inhomogeneities in the RF pulse and background field
  serve to effectively remove any signal perpendicular to
  the desired axis, and the induction signal is the same as in the first
  case.
Thus for example, if a measurement along $y$ for spin $I$ and along $x$ for
  spin $S$ were required, observing the induction signal after the sequence
\beqn
  \left[\frac{\pi}{2}\right]_{x}^{S}-\left[spin lock\right]_{x}^{I}.
\eeqn
Because one of the spins remains along the $z$--axis while the receiver is in
  phase with the other, the measured signals are anti-phase.
The spectrographic traces shown in Figs. 6a-d indicate the results of the
  measurements
  $\mbox{Tr}\left( 4I_xS_y \rho_{f}\right)$,
  $\mbox{Tr}\left( 4I_yS_x \rho_{f}\right)$,
  $\mbox{Tr}\left( 4I_yS_y \rho_{f}\right)$, and
  $\mbox{Tr}\left( 4I_xS_x \rho_{f}\right)$,
  respectively.
The traces show the Fourier-transformed induction signal read on the $^{13}$C 
channel, with absorptive peaks in phase along either the $+x$-- or $+y$--axis, 
depending on which axis the carbon nucleus was spin-locked.
Notice that Fig.\ \ref{eprfig}(d) shows the same anti-phase signal as the other
  spectra, but ``flipped'' by $180^{\circ}$.

The results of the four plots, taken together, show a simple inconsistency
compared to a model of only two uncorrelated classical magnetic dipoles.
The product of the four traces has an overall factor of 
$-1$, yet each magnetic moment is measured twice so that their signals should
cancel.  Each measurement is assumed to record either the x or y polarization
if each dipole is measured independently of the state of the other.

\section{Quantum Logic Gates}
NMR provides a means whereby it is possible to analyze experiments as building 
blocks for a quantum information processor (QIP).  Because spin $\tfrac{1}{2}$ 
particles can have two possible orientations (up or down), it is natural to 
associate spin states with computational bits.  Further, NMR experiments can
be viewed as performing computations on these quantum bits (qubits).  

\subsection{Pulse Sequences As Logic Gates}

Suppose we wanted to implement the controlled-NOT (c-NOT, or also XOR) gate, common in computer science, using NMR techniques.  A c-NOT gate performs a NOT operation on one bit, conditional on the other bit being set to 1.  The action of a c-NOT gate is summarized by the truth table
\begin{center}
\begin{tabular}{llll}
${\bf A_{input}}$ & ${\bf B_{input}}$ & ${\bf A_{output}}$ & ${\bf 
B_{output}}$\\
\cline{1-4}
F (up) & F (up) & F (up) & F (up)\\
F (up) & T (down) & F (up) & T (down)\\
T (down) & F (up) & T (down) & T (down)\\
T (down) & T (down) & T (down) & F (up),
\end{tabular}
\end{center}
where the True and False values have been associated with up spins and down
spins, respectively.  The above truth table corresponds to a unitary 
transformation that implements
\beqn
\begin{array}{rcl}
\mket{00} & \rightarrow & \mket{00}\\
\mket{01} & \rightarrow & \mket{01}\\
\mket{10} & \rightarrow & \mket{11}\\
\mket{11} & \rightarrow & \mket{10}.
\end{array}
\eeqn

In a weakly coupled two-spin system, a single transition can be excited via 
application of the propagator,
\beqn
{\cal U}\;=\;e^{-\imath \tfrac{1}{2} S_x \left(1 - 2 I_z \right) \omega t}\;=\;
\left(
\begin{array}{cccc}
1 & 0 & 0 & 0\\
0 & 1 & 0 & 0\\
0 & 0 & \cos{\tfrac{\omega t}{2}} & -\imath \sin{\tfrac{\omega t}{2}}\\
0 & 0 & \imath \sin{\tfrac{\omega t}{2}} & \cos{\tfrac{\omega t}{2}} 
\end{array}
\right),
\eeqn
which for a perfect $\omega t=\pi$ rotation becomes (to within a phase factor)
\beqn
{\cal U}\;=\;
\left(
\begin{array}{cccc}
1 & 0 & 0 & 0\\
0 & 1 & 0 & 0\\
0 & 0 & 0 & 1\\
0 & 0 & 1 & 0
\end{array}
\right).
\eeqn
It is clear that exciting a single transition in an NMR experiment is the same 
as a c-NOT operation from computer logic.   In NMR terms, the action of the c-NOT gate is to rotate one spin, conditional on the other spin being down.  Figure \ref{cnotfig} shows the result of performing a c-NOT on $\rho_{eq}$.
While NMR is certainly capable of implementing the c-NOT operation as is
done on a classical computer, that alone does not demonstrate any of the 
quantum dynamics.  Gates implemented on a quantum information processor 
which have no classical counterpart are of much more interest.  An example of 
such a gate is the single-spin Hadamard transform,
\beqn
H\;=\;\tfrac{1}{\sqrt{2}}
\left(
\begin{array}{cc}
1 & 1\\
1 & -1\\
\end{array}
\right)\;=\;e^{i\left(\tfrac{1}{2}-\tfrac{I_x+I_z}{\sqrt{2}}\right)\pi},
\label{hadamard}
\eeqn
which takes a spin from the state \ket{0} into the state $\tfrac{1}{\sqrt{2}} 
(\mket{0} + \mket{1})$.  This is just a $\pi$ rotation around the vector $45^o$ between the $x$ and $z$ axes.  A spectrum demonstrating the application of the Hadamard transform to the equilibrium state $\rho_{eq}$ is shown in figure \ref{Hadamardfig}.  The c-NOT and single-spin rotations can be combined to generate any desired unitary transformation, and for this reason they are referred to as a universal set of gates. \cite{BBCDMSSSW:95}


Analysis of conventional NMR experiments in terms of quantum information processing has led to a great deal of insight into areas such as the dynamics of pulse sequences for logic gates \cite{price}, and the effective Hamiltonian for exciting a single transition \cite{havel}.

\subsection{The Quantum Fourier Transform}
One of the most important transformations in quantum computing is the Quantum Fourier Transform (QFT).  The QFT is a necessary component of Shor's algorithm, which allows the factorization of numbers in polynomial time\cite{Shor}, a task which no classical computer can achieve (so far as is known).  Essentially, the QFT is the discrete Fourier transform which, for $q$ dimensions, is defined as follows 
\begin{equation}
QFT_q|a\rangle \rightarrow  \frac{1}{\sqrt{q}} \sum^{q-1}_{c=0} exp(2 \pi iac/q)|c\rangle
\end{equation}
This transform measures the input amplitudes of $|a\rangle$ in the $|c\rangle$ basis. Notice how the quantum Fourier transform on $|0\rangle$ will create an equal superposition in the $|c\rangle$ basis, allowing for parallel computation.  
In matrix form the two-qubit QFT transformation $QFT_2$, is expressed as 
\begin{eqnarray}
QFT_2 &=&
\frac{1}{2}
\left(
\begin{array}{cccc}
1 & 1 & 1 & 1 \\
1 & i & -1 & -i \\
1 & -1 & 1 & -1 \\
1 & -i & -1 & i \\
\end{array}
\right).
\end{eqnarray}
As formulated by Coppersmith \cite{cop}, the QFT can be constructed from two basic unitary operations; the Hadamard gate  $H_j$ (Eq.~\ref{hadamard}), operating on the {\it j}th qubit and the conditional phase transformation $B_{jk}$, acting on the {\it j}th and {\it k}th qubits, which is given by
\begin{eqnarray}
B_{jk} &=& 
\left( 
\begin{array}{cccc}
1 & 0 & 0 & 0 \\
0 & 1 & 0 & 0 \\
0 & 0 & 1 & 0 \\
0 & 0 & 0 & e^{{i\theta_{jk}}}
\end{array}
\right)
\;=\; e^{i\theta_{jk}\frac{1}{2}(1-2I_z)\frac{1}{2}(1-2S_z)}
\label{Bgate}
\end{eqnarray}  
where $\theta_{jk} = \frac{\pi}{2^{k-j}}$.  The two-qubit QFT, in particular, can be constructed as
\begin{equation}
QFT_2\;=\; H_0B_{01}H_1
\end {equation}
The $B_{jk}$ transformation can be implemented by performing the chemical shift and coupling transformations shown in Eq.~\ref{Bgate}. Figure \ref{QFTfig} shows the implementation of the QFT on a two spin system. The spectra show the 90$^{\rm o}$ phase shifts created after the QFT application.

\section{Conclusion}
Several basic but important concepts relevant to QIP are illustrated by experiments on a liquid-state ensemble NMR quantum information processor.  While pure quantum mechanical states are not achievable here, the creation and application of pseudo-pure states is demonstrated.  Tests of spinor behavior and entanglement are also described, illustrating quantum mechanical dynamics.  Finally, building blocks (the Hadamard, c-NOT, and QFT) for a more complicated quantum computer are also introduced.

\bibliography{concepts3}
\bibliographystyle{unsrt}

\newpage
Below are the Captions for the Figures.

\begin{enumerate}
\item At room temperature, the equilibrium state of chloroform molecules in solution is described by $\gamma_II_z+\gamma_SS_z$.  In order to create a pseudo-pure state from the equilibrium state, it is convenient to start with an equalized magnetization for I and S.  Since the ratio of $\gamma_I$ to $\gamma_S$ is a factor of four, then the spectra of I and S following a $\tfrac{\pi}{2}$ pulse should reflect the 4:1 ratio in the peak heights (figure (a)).  In order to compensate for the different electronics in the two channels, the gains of the channels were manually calibrated to produce the desired 4:1 ratio in signal intensity.  After this, the pulse sequence discussed in the text was applied.  A subsequent $\tfrac{\pi}{2}$ read out pulse results in the spectrum of figure (b).  The peaks have equal intensity, confirming the creation of the state $\tfrac{\gamma_I+\gamma_S}{2}(I_z+S_z)$.
\label{ppprepfig}

\item Once the pseudo-pure state $\rho_{pp}=I_z+S_z+2I_zS_z$ has been prepared, we use readout pulses to generate a series of spectra confirming that the desired state has been created.  This is done by applying $\tfrac{\pi}{2}|^S_y$, $\tfrac{\pi}{2}|^I_y$, and $\tfrac{\pi}{2}|^{I,S}_y$ read pulses on the pseudo-pure state.  The results are shown in figures (a)-(c), respectively, on both the carbon and hydrogen channels. The signature of the appropriate terms in $\rho_{pp}$ is seen from the three sets of spectra generated.
\label{pptomofig}

\item The state $2I_zS_x$ correlates the spinor behavior of spin I to the reference spin S.  The propagator $U=e^{-i\phi I_y(\tfrac{1}{2}-S_z)}$ then rotates all the I spins coupled to the down S spins by the angle $\phi$ about the y-axis.  Applying $U$ to the density matrix $2I_zS_x$ creates the state $2\cos(\phi/2)I_zS_z+2\sin(\phi/2)I_xS_x$, where only the first (antiphase) state is made observable by evolution under the internal Hamiltonian.  When $\phi=0$, the state is of course $2I_zS_x$, as shown in figure (a).  When $\phi=2\pi$, this state is inverted, contrary to common intuition.  The resulting spectrum is shown in figure (b).  Only when $\phi=4\pi$ does the antiphase state return to its original state as seen in the spectrum (c).  These spectra clearly demonstrate the spinor behaviour of spin $\tfrac{1}{2}$.
\label{spinorfig}

\item The pseudo-pure Bell state, $\rho_{Bell}=2I_zS_z+2I_xS_x-2I_yS_y$, created by the application of the propagator, $U=e^{iI_xS_y\pi}$ on the pseudo-pure state discussed above can be verified by applying a series of readout pulses on $\rho_{Bell}$.  Using the read pulses $\tfrac{\pi}{2}|^S_y$, $\tfrac{\pi}{2}|^S_x$, $\tfrac{\pi}{2}|^I_y$, and $\tfrac{\pi}{2}|^I_x$ on $\rho_{Bell}$, figures (a)-(d), respectively, and observing the resulting spectra on both the I and S channels confirms both the signature and the individual terms of $\rho_{Bell}$.
\label{belltomofig}

\item We simulate a strong measurement (one that collapses the wave function
along a preferred basis or axis) on a Bell State using magnetic field gradients.  An x-measurement on the I-spin is imitated by applying a selective 
x-gradient to it.  Since the two spin state is entangled, this measurement necessarily collapses the S-spin along the x-direction.  Subsequent measurements confirm that both I and S spins have transformed identically and that they are aligned along the x-axis.  This was verified by observing the creation of the $2I_xS_x$ state where in (a) we observe immediately after the ``strong measurement'' in both channels and see zero signal as expected.  In (b) we show that $\tfrac{\pi}{2}$ pulses along the x-axis has no effect and in (c) we verify that a $\tfrac{\pi}{2}|^S_y$  pulse indeed creates an anti-phase signal on the carbon channel and a $\tfrac{\pi}{2}|^I_y$ pulse creates an antiphase signal on the hydrogen channel.
\label{strongmeasure}


\item
Strong Measurements After EPR Preparation.  All four measurements are made on
the Carbon (1st spin) channel, and show the expected anti-phase correlation.
({\em a}) Measurement of correlation $I_yS_x$, in phase along the $+y$
direction. ({\em b}) Measurement of correlation $I_xS_y$, in phase along
the $+x$ direction. ({\em c}) Measurement of correlation $I_yS_y$, in
phase along the $+y$ direction. ({\em d}) Measurement of correlation
$I_xS_x$, in phase along the $+x$ direction. Note that the last spectrum is
``flipped,'' or inverted, with respect to the other three.  
\label{eprfig}

\item The above spectra show the implementation of a controlled-NOT (c-NOT) gate on the equilibrium state of $^{13}$C-chloroform.  The spectrum on the right represents the readout on the I spins, and the spectrum on the left is the readout on the S spins.  Both spectra have the expected appearance and confirm the creation of the state $I_z+2I_zS_z$, the expected state after application of the c-NOT.
\label{cnotfig}

\item
The Hadamard gate $H$ is a one bit gate that can be geometrically interpreted as a $\pi$ rotation about the $\tfrac{1}{\sqrt{2}}(x+z)$ axis.  If the net magnetization is along the $+y$ direction then the Hadamard gate should simply rotate it to the $-y$ direction (figure (a)).  However, since any $\pi$ rotation about an axis in the x-z plane performs the same transformation, $H$ was also applied to an initial $+x$ magnetization.  The result (figure (b)) shows how the magnetization was sent to the z-axis, as expected.  
\label{Hadamardfig}

\item The two-qubit QFT was implemented by applying a Hadamard gate on the first spin, a conditional phase operator, and a Hadamard on the second spin.  A Hadamard gate can be performed by a simple combination of three pulses: $\tfrac{\pi}{4}|_x - \pi|_y - \tfrac{\pi}{4}|_{-x}$.  Because it was performed on the thermal state, the initial Hadamard was simplified to a $\pi/2_y$ pulse.  The conditional phase change operator, $B_{01}$, was implemented by delay 2 and pulses 3 to 6, where pulses 4 to 6 are a $\pi/4$ z-rotation.  The final Hadamard gate was implemented by the three pulses labeled 7.  The phase difference of each peak on the spectra shows the two-bit QFT's ability to separate input states by 90 degrees.  After the application of the QFT, the spins were phase shifted by 45 degrees and were allowed to evolve for a time 1/4J in order to bring out the phase differences.  
\label{QFTfig}

\end{enumerate}

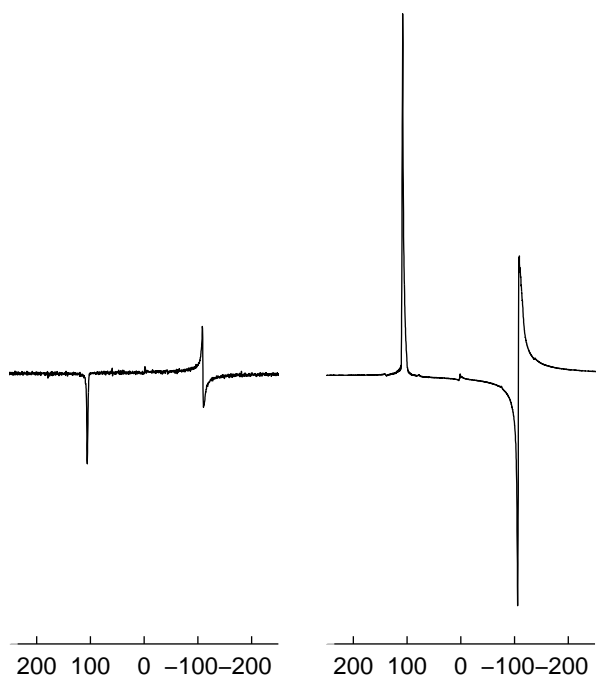
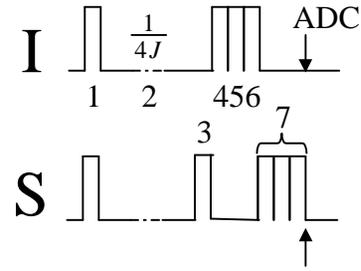
\end{document}